\documentclass[twocolumn,showpacs,preprintnumbers,amsmath,amssymb]{revtex4}
\usepackage{amsfonts}
\usepackage{amsmath}
\usepackage{bbm}
\usepackage{graphicx}
\usepackage{bm}
\usepackage{dcolumn}

\begin{document}


\title{Photonic quantum memory in two-level ensembles based on modulating the refractive index in time: equivalence to gradient echo memory}

\author{James Clark, Khabat Heshami and Christoph Simon}
\affiliation{Institute for Quantum Information Science
and Department of Physics and Astronomy, University of
Calgary, Calgary T2N 1N4, Alberta, Canada}

\date{\today}

\begin{abstract}
We present a quantum memory protocol that allows to store light in ensembles of  two-level atoms, e.g. rare-earth ions doped into a crystal, by modulating the refractive index of the host medium of the atoms linearly in time. We show that under certain conditions the resulting dynamics is equivalent to that underlying the gradient echo memory protocol, which relies on a spatial gradient of the atomic resonance frequencies. We discuss the prospects for an experimental implementation.
  \end{abstract}

\maketitle
\section{\label{Intro}Introduction}
Quantum memory for light is an essential element for the photonic implementation of quantum communication and information processing \cite{Kok07,SanguardRMP}. In recent years there has been a lot of work both on theoretical proposals and on experimental implementations \cite{Lvovsky09,Simon10}. Thus far optical control, using relatively strong laser pulses, has been exploited for electromagnetically induced transparency and off-resonant Raman-type storage \cite{Phillips01,Reim10} in ensembles of three-level systems. More recently, a direct control of the transition dipole-moment has been proposed that emulates Raman-type quantum memories in a two-level atomic configuration \cite{CDmemory}. 

In photon-echo based memories \cite{Tittelreview}, the light-matter coupling is controlled in a more indirect way by exploiting the dephasing and rephasing of inhomogeneously broadened atomic ensembles. This includes the controlled reversible inhomogeneous broadening protocol \cite{CRIB}, the atomic frequency comb protocol \cite{AFC}, and the gradient echo memory (GEM) protocol \cite{GEM}. The GEM protocol has allowed the demonstration of the highest memory efficiency (in the quantum regime) so far \cite{Hedges10}.

Recently Ref. \cite{Kalachev} proposed a quantum memory protocol based on controlling the refractive index. Considering an ensemble of three-level atoms inside a host medium (e.g. rare-earth ions doped into a crystal), which is located in a circular optical cavity, they showed that a continuous change of the refractive index of the host medium during an off-resonant Raman interaction between a single photon, a classical control pulse and the atomic ensemble allowed mapping the state of the single photon into a collective atomic excitation. 

 Here we consider storing quantum states of light in an ensemble of two-level atoms in a host medium, where the refractive index of the medium can be modulated during the interaction of the light with the atoms. In contrast to Ref. \cite{Kalachev} here there is no optical control pulse (which is related to the fact that we consider two-level instead of three-level atoms) and no cavity. Interestingly, we find that under certain conditions the considered system leads to dynamics that are equivalent to those of the GEM protocol; controlling the refractive index of the host medium {\it in time} can mimic the effect of the {\it spatial} frequency gradient present in GEM.

This paper is organized as follows. In Sec. \ref{Max-Bloch} we derive the dynamical equations for our system under certain conditions, which we discuss in detail. In Sec. \ref{Comparison} we compare these results to the GEM protocol. In Sec. \ref{Implementation} we discuss a possible experimental implementation of our protocol. Sec. V contains our conclusions.

\section{\label{Max-Bloch}Maxwell-Bloch equations}

Here we study the propagation of the light and its interaction with two-level atoms inside a host medium whose refractive index varies in time. We show that in a certain parameter regime the time-dependent refractive index does not play a role in the propagation equation for the light. In contrast, it plays an essential role in the dynamics of the atomic polarization. For simplicity, we assume the field is propagating in a certain direction with a fixed linear polarization. (This is well justified for our choice of possible implementation in a waveguide, see below.) The wave equation for the electric field operator is analogous to the classical equation, namely
\begin{equation}\label{wave1}
\frac{\partial^2E}{\partial z^2} =\mu_0\frac{\partial^2 D} {\partial t^2} = \mu_0\frac{\partial^2}{\partial t^2}(\epsilon E+P),
\end{equation}
where $E$ is the electric field, $z$ is the direction of propagation, $\mu_0$ is the vacuum permeability, $D$ is the electric displacement field, $\epsilon$ is the permittivity of the propagation medium and $P$ is the polarization of the embedded two-level atoms. There are thus two fundamentally different contributions to $D$. The $\epsilon E$ term is due to the permittivity of the host medium, whereas $P$ describes the polarization of the two-level atoms that are the actual memory system for the light.

Consider the case where $\epsilon$ is time-dependent. The permittivity of the medium is related to its refractive index as $\epsilon(t)=n^2(t) \epsilon_0$. We consider a medium with a linearly changing refractive index, $n(t)=n_i+{\dot n}t$. Based on this, only the first derivative of the refractive index remains in the Eq. (\ref{wave1}), giving
\begin{equation}\label{wave2}
(\frac{\partial^2}{\partial z^2} -\frac{n^2(t)}{c^2} \frac{\partial^2}{\partial t^2})E = \frac{1}{c^2}(2{\dot n}^2E+4n(t){\dot n}{\dot E})+\mu_0 \ddot{P},
\end{equation}
where $c$ is the speed of light. We now introduce the slowly varying components of the signal field $E$ and the atomic polarization $P$, $E={\cal E}e^{-i(\omega_0t-k_0(t)z)}$ and $P={\cal P}e^{-i(\omega_0 t-k_0(t)z)}$. Here the wave vector $k_0(t)=k_i+{\dot k}t=(n_i+{\dot n}t)\frac{\omega_0}{c}$ is a function of time and ${\dot k}=\frac{{\dot n}\omega_0}{c}$, where $\omega_0$ is the central frequency of the signal.

The wave equation can be greatly simplified provided that a number of (realistic) conditions are fulfilled (see also the appendix). The second-order spatial derivative for ${\cal E}$ can be dropped provided that the field amplitude changes appreciably over the length of the medium (such that the derivative is comparable to ${\cal E}/L$) and
 $k_0(t)\gg 1/L$. Similarly, the second order time-derivative can be ignored if $\omega_0\gg 1/{\tau}$, where $\tau$ is the duration of the pulse. The same conditions also allow one to drop the first and second order derivatives of the slowly varying polarization operator. We are interested in the regime where the extent of the pulse in space (outside the medium), ${\cal L}=c\tau$, is much greater than the length of the medium, $L$ \cite{Longdell08,AFC}. This allows one to drop the first-order time derivative of ${\cal E}$ compared to the first-order spatial derivative. Finally, by assuming $\Delta n \ll n_i$, where $\Delta n$ is the total change in the refractive index, and ${\dot k} L\ll \frac{2c}{nL}$ one obtains the simplified propagation equation
   \begin{equation}\label{wave-slow2}
\frac{\partial {\cal E}}{\partial z}=\frac{i\mu_0\omega_0^2}{2k_i}{\cal P}.
\end{equation}
This shows that, under the above conditions, the propagation equation remains unchanged compared to that of systems with constant refractive index (${\dot k}$ does not play an appreciable role in the propagation), see e.g. \cite{Longdell08,AFC}. The derivation of Eq. (3) from Eq. (2) is discussed in detail in the appendix.

We now derive the dynamics of the polarization of the dopant atoms.
The polarization of the $j$-th atom is $P^{j}=\langle g^{j}| {\hat d} |e^{j}\rangle\sigma_{ge}^{j}$, where $\sigma_{ge}^j=|g^j\rangle\langle e^j|$ and $\langle g^{j}| {\hat d}|e^{j}\rangle$ is the matrix element of the corresponding dipole moment component between the ground and excited states. The collective atomic polarization at a certain position $z$ is the sum over the individual atoms in a slice of width $\Delta z$. The slow component of this collective operator is given by
\begin{equation}\label{pol-def}
{\cal P}=\frac{1}{A\Delta z} \langle g|  {\hat d} |e\rangle \sum_{j=1}^{N_z} \sigma_{ge}^{j}e^{i(\omega_0 t - k_0(t)z_j)}\equiv \langle g| {\hat d}|e\rangle \frac{N}{V} {\tilde \sigma}_{ge},
\end{equation}
where we assume equivalent dipole moment for all of the atoms; $A$ and $V$ are the cross-section area and volume of the light-atom interface; $N$ is the number of the dopant atoms and ${\tilde \sigma}_{ge}=\frac{1}{N_z}\sum_{j=1}^{N_z}\sigma_{ge}^j e^{i{(\omega_0 t-k_0(t)z)}}$ is the average atomic polarization at position $z$.
The Hamiltonian of the ensemble of the dopant atoms interacting with the light field can be written as
\begin{eqnarray}\label{hamiltonian1}
H&=&H_0+H_{int}\\ \nonumber
&=&\sum_{j=1}^{N}\hbar\Omega\sigma_{ee}^j - \langle e|{\hat d}| g\rangle \sum_{j=1}^{N} \sigma_{eg}^j E(z_j,t)+ h.c.,
\end{eqnarray}
where we assume uniform excited state energy $\hbar \Omega$ for all of the atoms. 
We can now derive the dynamics of the slowly varying collective atomic polarization using
\begin{equation}\label{heisenberg}
\frac{d\tilde{\sigma}_{ge}}{dt}= -\frac{i}{\hbar}[{\tilde \sigma_{ge}},H]+ \frac{\partial {\tilde \sigma_{ge}}}{\partial t}.
\end{equation}
For the next step we assume that all of the atoms are initialized in the ground state and the number of atoms $N\gg 1$. For weak (quantum) signals one can then ignore the change in the excited state population. 

Using Eqs. (\ref{wave-slow2},\ref{pol-def},\ref{hamiltonian1},\ref{heisenberg}), the above definition of the slowly varying field,
and including the atomic excited state linewidth $\gamma$, 
one finds the Maxwell-Bloch equations describing the interaction of the light with the collective atomic polarization in a medium with linearly time-varying refractive index,
\begin{eqnarray}\label{Max-Blocheq}
\frac{d{\tilde \sigma_{ge}}(z,t)}{dt} &=& -(\gamma+i(\Delta+{\dot k}z)){\tilde \sigma_{ge}}(z,t)+ig{\tilde{\cal E}}(z,t),\\ \nonumber
\frac{\partial{\tilde{\cal E}}(z,t)}{\partial z}&=& i\frac{nNg}{c} {\tilde \sigma}_{ge}(z,t),
\end{eqnarray}
where ${\tilde{\cal E}}=\sqrt{\frac{\hbar \omega_0}{2\epsilon_i V}}{\cal E}$ and $\Delta=\Omega -\omega_0$ is the detuning. The coupling constant $g=\langle e| {\hat d}|g\rangle \sqrt{\frac{\omega_0}{2\hbar\epsilon_i V}}$, where $\omega_0$ is the central frequency of the pulse and $\epsilon_i=n_i^2\epsilon_0$ is the initial refractive index of the medium. Note that the time dependence of the permittivity can be ignored in the definitions of $\tilde{{\cal E}}$ and $g$ because we are interested in the regime where $\Delta n \ll n_i$.

\begin{figure}[t]
\scalebox{0.45}{\includegraphics{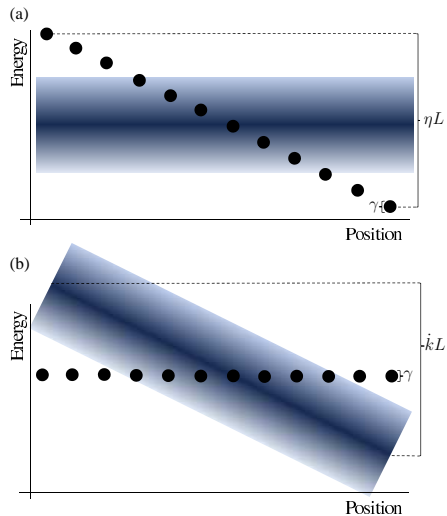}}
\caption{\label{fig1} (a) In the GEM protocol, a longitudinal energy shift in the atoms (solid dots) allows one to cover all of the frequency components of the incoming light.
(b) In the protocol proposed here, due to the linear change of the refractive index in time, the light experiences an effective position-dependent frequency shift ${\dot k}z$. This allows different frequency components of the light to interact on resonance with a spectrally narrow line of atoms.}
\end{figure}

The above set of equations shows the role of the linearly changing refractive index of the host medium in the regime that we have discussed. One sees that the linear change of the refractive index in time results in a space-dependent frequency shift given by the ${\dot kz}$ term in Eq. (\ref{Max-Blocheq}), see also Fig. \ref{fig1}(b). The above Maxwell-Bloch equations are identical to those underlying the GEM quantum memory protocol \cite{GEM}. In the next section we therefore discuss in detail how the present proposal compares to GEM.

\section{\label{Comparison} Comparison with Gradient Echo Memory}

The dynamical equations (8) derived in the previous section (under a number of realistic conditions) are exactly equivalent to those underlying the GEM quantum memory protocol \cite{GEM}. In the GEM protocol, an initially narrow atomic absorption line is broadened by applying an external (longitudinal) field gradient. This longitudinal broadening allows one to accommodate all frequency components of the incoming pulse, see Fig. \ref{fig1}(a). Once the pulse is absorbed, the produced collective atomic excitation starts to dephase due to the position-dependent detuning. By inverting the external field that is used to generate the gradient, one can rephase the collective atomic excitation, which leads to the re-emission of the light.

At first sight it may seem surprising that a time variation of the refractive index leads to a spatial gradient in the detuning. This can be understood in the following way. In the definition for the slow-varying field ${\cal E}$, $E={\cal E}e^{-i(\omega_0t-k_0(t)z)}$, the fast-varying phase $-i(\omega_0t-k_0(t)z)$ has a temporal and a spatial part. One can define an effective frequency for the light by taking the time derivative of the phase, $\omega_{eff}=\omega_0 - {\dot k} z$. One can see that this corresponds to a spatial gradient in the frequency of the light, leading to a spatial gradient in the light-atom detuning, see Fig. \ref{fig1}(b). A quantum memory can then be realized in analogy with GEM. The light is absorbed while the refractive index is changing linearly in time. Then the refractive index is kept constant for storage. The light can be retrieved by changing the refractive index linearly in time again, but with the opposite sign to before. As for other two-level memory protocols including GEM, the storage time can be increased if necessary by transferring the population (after the absorption has been completed) from the excited state to a long-lived third state.

The total shift in the effective frequency over the length of the medium is ${\dot k} L$. In order to accommodate all frequency components of the signal ${\dot k} L$ has to be larger than the frequency bandwidth of the signal, $\Delta \omega \equiv 1/\tau$.
Therefore, ${\dot k} L$ can be understood as the memory bandwidth.
This is equivalent to the role of $\eta L$ in GEM, see Fig. \ref{fig1} and \cite{Longdell08}.

\begin{figure}
\scalebox{0.46}{\includegraphics{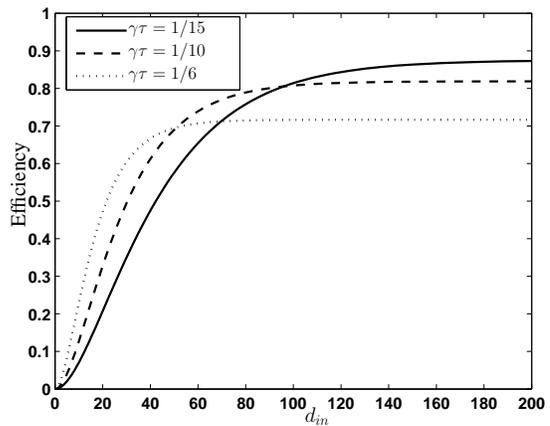}}
\caption{\label{fig2} The efficiency of the proposed memory protocol based on refractive index modulation in terms of the initial optical depth $d_{in}$. The efficiency is given by $e^{-2\gamma \tau}(1-\exp{(-d_{in}\gamma/{\dot k}L)})^2$. The figure shows the efficiency for different pulse durations $\tau$, relative to the excited state line width, $\gamma$. We assume ${\dot k}L \tau=2$. Depending on the available optical depth, one can optimize the achievable efficiency by choosing an appropriate pulse duration.}
\end{figure}

The efficiency of the present quantum memory proposal can be found by analogy to GEM, see  Eqs. (3, 4) in \cite{Longdell08}. Converting the equations of motion to the frequency domain, one finds that the transmitted pulse is attenuated by a factor of $\exp(-\beta \pi)$, where $\beta=\frac{nNg^2}{c{\dot k}}$. This implies that the optical depth of the system is $d=2\beta \pi=2\pi \frac{nNg^2}{c{\dot k}}=d_{in}\frac{\gamma}{{\dot k}L}$. Here $d$ is the optical depth that is associated with the effectively broadened line, with the initial optical depth $d_{in} \equiv 2\pi \frac{nNg^2L}{c\gamma}$.
The retrieval efficiency is then given by $(1-\exp{(-d)})^2 e^{-2\gamma\tau}$, see Fig. \ref{fig2}. Here we have assumed that the decay of the excited state only has an effect during absorption and retrieval, but not during storage. As mentioned before, this can be achieved e.g. by transferring the excitation to a third, longer-lived state. Hyperfine ground states in rare-earth doped crystals can have coherence times of many seconds \cite{Fraval}.

\section{\label{Implementation} Possible implementation}

We now discuss a potential experimental implementation of our proposed protocol. 
We propose to use thulium ions doped into a lithium niobate waveguide. This system was used in a recent implementation of an atomic frequency comb memory \cite{Erhan11}. Lithium niobate is an attractive host for the present proposal because of its electro-optic properties, see below. The thulium ions interact with near-infrared light at a wavelength of 795 nm. The transition is naturally inhomogeneously broadened. We propose to prepare an initial atomic linewidth of $\gamma=10$ MHz by optical pumping, which is very realistic. We consider the case where ${\dot k}L\tau = 2$. This assures that the memory bandwidth is large enough that it can accommodate the incoming pulse. Assuming $L=1$ cm this leads to the requirement $\Delta n \approx 5*10^{-5}$. We choose the pulse duration $\tau=1/10\gamma$. The above values assure that $\omega_0\gg 1/\tau$, $L\ll {\cal L}=c\tau$, ${\dot k}L\ll \frac{2c}{nL}$, $\Delta n \ll n_i$ and $k_0(t)\gg 1/L$, as required for the derivation in section II. For the given parameter values one needs $d_{in}=50$ to achieve about 70\% efficiency, see also Fig. \ref{fig2}.

We consider the refractive-index modulation of the ordinary optical axis of lithium niobate by a fast varying electric field. We consider the case where the crystal is clamped (spatially confined) and the temperature is a few Kelvin. Under these conditions the refractive index of the ordinary axis $n_o \approx 2.26$ \cite{Wongbook}. The change in the refractive index through the electro-optical effect is governed by $\Delta (\frac{1}{n^2})_{ij} =\sum_{k}r_{ijk}E_k$, where $i,j=1$ is associated with the refractive index change of the ordinary axis. This means that a time-dependent external field in a certain direction, $E_k$, can impose a time-dependent refractive index for the ordinary axis if there exists a non-zero linear electro-optical coefficient for that direction, $r_{11k}$. For lithium niobate $r_{113}\approx 10*10^{-12}$m/V and $r_{112}\approx -3*10^{-12}$m/V \cite{Wongbook, Weis85}.
This leads to $\Delta n\approx 6*10^{-5}$ under $0.85*10^{6}-2.5*10^{6}$V/m electric field, depending on the direction of the field in the 2-3 plane. This is equivalent to applying $8.5-25$V to a system that has $10\mu$m thickness, comparable to the waveguide used in Ref. \cite{Erhan11}. The maximum change in the refractive index in lithium niobate is expected to be $10^{-3}$, which is limited by the breakdown electric field \cite{Kalachevimp}.

Applying the external electric field to change the refractive index is potentially accompanied by level shifts, due to the linear Stark shift, for the atomic ground and excited states. On the other hand, for a certain type of dopant, by having the external electric field orthogonal to the difference between the permanent electric dipole moment of the ground and excited states, one can keep the resonant frequency between these states unchanged. In our proposed system the permanent dipoles are aligned with the 3-axis \cite{Sinclair09,Thiel}, therefore the electric field should be applied along the 2-axis in order to avoid level shifts.

\section{\label{Conclusion}Conclusion}

We have proposed a memory protocol based on varying the refractive index of the host medium in time, and shown that in a certain regime it is equivalent to the GEM protocol, even though the latter is based on a spatial frequency gradient. One may wonder why no similar spatial gradient was seen in the protocol of Ref. \cite{Kalachev}, which also considered a time-varying refractive index, but in a Raman-type system, in contrast to the two-level ensemble considered by us. This can be understood by noting that in the scheme of Ref. \cite{Kalachev} the refractive index modulation causes a spatial gradient in the frequencies for both the signal and the control fields, such that the two-photon transition frequency remains unchanged. This holds for co-propagating signal and control. It might be interesting to consider counter-propagating signal and control fields in this context, for which no such cancelation would occur. This might result in a protocol similar to Raman-GEM \cite{Raman-GEM}.

We found that a relatively small modulation of the refractive index should be able to provide sufficient memory bandwidth. We proposed a potential implementation in lithium niobate waveguides doped with rare-earth ions. However, other implementations could also be considered. In general, it may be easier to control the behavior of the refractive index in time, compared to the spatial control of the atomic transition frequencies required in standard GEM experiments.

{\it Acknowledgments.} This work was supported by NSERC and AITF. We thank H. Kaviani, E. Saglamyurek and W. Tittel for fruitful discussions.

\section{\label{Appendix} Appendix: Wave equation simplification}

Here we explain in more detail how to obtain Eq. (3) from Eq. (2). Substituting $E={\cal E}e^{-i(\omega_0t-k_0(t)z)}$ and $P={\cal P}e^{-i(\omega_0 t-k_0(t)z)}$ into Eq. (\ref{wave2}) and
canceling the fast varying phase gives
\begin{eqnarray}
&&(\partial_z^2 + 2ik_0(t)\partial_z - k_0^2(t)){\cal E}\\ \nonumber && -\frac{n^2(t)}{c^2} (\partial_t^2 - 2i (\omega_0 - {\dot k}z)\partial_t - (\omega_0 - {\dot k}z)^2){\cal E} \\ \nonumber && =
\frac{1}{c^2}(2{\dot n}^2 - 4in(t){\dot n}(\omega_0 -{\dot k}z)+ 4n(t){\dot n}\partial_t){\cal E}  \\ \nonumber && + \mu_0(\partial_t^2 - 2i(\omega_0 -{\dot k}z)\partial_t - (\omega_0 -{\dot k}z)^2){\cal P}.
\end{eqnarray}
In the left hand side of the equation, the term $k_0^2(t){\cal E}$ cancels $\frac{n^2(t)}{c^2}\omega_0^2{\cal E}$. The second order spatial derivative of the field can be dropped compared to the first order, assuming that the field changes appreciably over the length of the medium, $L$, and $k_0(t)\gg \frac{1}{L}$. Similarly, the second order time derivative of the field and the first and second order time derivative of the polarization operator can be ignored compared to the first order derivatives if $\omega_0\gg \frac{1}{\tau}$. This simplification is valid as long as $\omega_0\gg {\dot k}L$. These conditions lead to
\begin{eqnarray}
&&(2ik_0(t)\partial_z+\frac{n^2(t)}{c^2} ( 2i \omega_0 \partial_t -2 \omega_0{\dot k}z )){\cal E} \\ \nonumber && =
\frac{1}{c^2}(2{\dot n}^2 - 4in(t){\dot n} \omega_0+ 4n(t){\dot n}\partial_t){\cal E} - \mu_0 \omega_0^2 {\cal P}.
\end{eqnarray}
One can rewrite this equation and use $k_0(t)=n(t)\omega_0/c$ to simplify it to
\begin{eqnarray}
&&(\partial_z+\frac{n(t)}{c}(\partial_t +i{\dot k}z )){\cal E} \\ \nonumber &&  = (-\frac{i{\dot n}^2}{c^2 k_0(t)} - \frac{2{\dot n}}{c} - \frac{2in(t){\dot n}}{c^2 k_0(t)} \partial_t){\cal E} + \frac{i\mu_0 \omega_0^2}{2k_0(t)} {\cal P}.
\end{eqnarray}
 Provided that $\Delta n \ll n(t)$ (and using again $\omega_0\gg 1/\tau$), one has
  $\frac{i{\dot n}^2}{c^2 k_0(t)}{\cal E}\ll \frac{2in(t){\dot n}}{c^2 k_0(t)} \partial_t{\cal E} \ll i\frac{n(t){\dot k}z}{c} {\cal E}$. The condition $\Delta n \ll n(t)$ also allows one to drop $\frac{2{\dot n}}{c}{\cal E}$ compared to $\frac{n(t)}{c}\partial_t {\cal E}$. This simplifies the wave equation to
\begin{equation}
(\partial_z+\frac{n(t)}{c}(\partial_t +i{\dot k}z )){\cal E} = \frac{i\mu_0 \omega_0^2}{2k_0(t)} {\cal P}.
\end{equation}
Provided that the extent of the pulse in space, ${\cal L}=c\tau$, is much larger than the length of the medium, $L$, one can ignore $\frac{n(t)}{c}\partial_t {\cal E}$ in comparison with $\partial_z {\cal E}$, which is the dominant term for the slowly varying component of the field. Finally, assuming that ${\dot k}L\ll \frac{2c}{nL}$ the term $\frac{in(t){\dot k}z}{c}{\cal E}$ also can be dropped compared to the dominant term, leading to
\begin{equation}
\partial_z {\cal E} = \frac{i\mu_0 \omega_0^2}{2k_0(t)} {\cal P}.
\end{equation}
The coefficient of ${\cal P}$ can be rewritten as,
\begin{equation}
\frac{i\mu_0 \omega_0^2}{2k_i(1+\frac{{\dot k}t}{k_i})}\approx \frac{i\mu_0 \omega_0^2}{2k_i}(1-\frac{{\dot k}t}{k_i}).
\end{equation}
 Under the condition $\Delta n \ll n_i$ this can be well approximated by $\frac{i\mu_0 \omega_0^2}{2k_i}$, which leads to
\begin{equation}
\partial_z {\cal E} = \frac{i\mu_0 \omega_0^2}{2k_i} {\cal P},
\end{equation}
which is the standard wave equation under similar (realistic) conditions in the absence of a time variation of the refractive index.


\begin{thebibliography}{99}


\bibitem{Kok07} P. Kok {\it et al.}, Rev. Mod. Phys. {\bf 79}, 135 (2007).

\bibitem{SanguardRMP} N. Sangouard, C. Simon, H. de Riedmatten, and N. Gisin, Rev. Mod. Phys. {\bf 83}, 33 (2011).


\bibitem{Lvovsky09} A.I. Lvovsky, B.C. Sanders and W. Tittel, Nature Photonics {\bf 3}, 706 (2009).

\bibitem{Simon10} C. Simon {\it et al.}, Eur. Phys. J. D {\bf 58}, 1 (2010).

\bibitem{Phillips01} D.F. Phillips, A. Fleischhauer, A. Mair, and R.L. Walsworth, Phys. Rev. Lett. {\bf 86}, 783 (2001).

\bibitem{Reim10} K.F. Reim {\it et al.}, Nature Photonics {\bf 4}, 218 (2010).

\bibitem{CDmemory} A. Green {\it et al.}, arXiv:1106.3513 (2011).

\bibitem{Tittelreview} W. Tittel {\it et al.}, Laser Photonics Reviews {\bf 4}, 244 (2009).
    

\bibitem{CRIB} S.A. Moiseev S. Kr\"{o}ll, Phys. Rev. Lett. {\bf 87}, 173601 (2001); B. Kraus {\it et al.}, Phys. Rev. A {\bf 73}, 020302(R) (2006).

\bibitem{AFC} M. Afzelius, C. Simon, H. de Riedmatten, and N. Gisin, Phys. Rev. A {\bf 79}, 052329 (2009).


\bibitem{GEM} G. H\'{e}tet, J.J. Longdell, A.L. Alexander, P.K. Lam, and M.J. Sellars, Phys. Rev. Lett. {\bf 100}, 023601 (2008).

\bibitem{Hedges10} M.P. Hedges, J.J. Longdell, Y. Li and M.J. Sellars    Nature {\bf 465}, 1052 (2010).

\bibitem{Kalachev} A. Kalachev and O. Kocharovskaya, Phys. Rev. A {\bf 83}, 053849 (2011).

\bibitem{Longdell08} J.J. Longdell, G. H{\' e}tet and P.K. Lam, M.J. Sellars Phys. Rev. A {\bf 78}, 032337 (2008).

\bibitem{Fraval} E. Fraval, M.J. Sellars, and J.J. Longdell, Phys. Rev. Lett. {\bf 95}, 030506 (2005).


\bibitem{Erhan11} E. Saglamyurek {\it et al.}, Nature (London) {\bf 469}, 512 (2011).

\bibitem{Wongbook} K.K. Wong, {\it Properties of Lithium Niobate} (INSPEC, London, 2002).

\bibitem{Weis85} R.S. Weis and T.K. Gaylord, Appl. Phys. A {\bf 37}, 191 (1985).

\bibitem{Kalachevimp} A. Kalachev and O. Kocharovskaya, J. Mod. Opt. {\bf 58}, 1971(2011).

\bibitem{Sinclair09} N. Sinclair {\it et al.}, Journal of Luminescence {\bf 130} 1586 (2010).

\bibitem{Thiel} C.W. Thiel, Y. Sun, T. B{\" o}ttger, W.R. Babbitt and R.L. Cone, Journal of Luminescence {\bf 130} 1598 (2010).
    
\bibitem{Raman-GEM} M. Hosseini {\it et al.}, Nature {\bf 461}, 241 (2009).

\end{thebibliography}
\end{document}